\documentstyle[pra,multicol,aps]{revtex}
\begin{document}
\setlength{\textwidth}{150mm}
\setlength{\textheight}{240mm}
\setlength{\parskip}{1mm}
\input{epsf.tex}
\epsfverbosetrue

\title{\bf Quadratic Solitary Waves in a
Counterpropagating Quasi-Phase-Matched 
Configuration}

\author{Kazimir Y.~Kolossovski \and Alexander V.~Buryak \and 
Rowland A.~Sammut \\ 
\protect\small\textit{School of Mathematics and Statistics, 
University College, Australian Defence Force Academy, 
Canberra 2600, Australia}}

\maketitle

\begin{abstract}
We demonstrate the possibility of self-trapping of optical beams 
by use of quasi phase matching in a counterpropagating configuration
in quadratic media. We also show the predominant stability of 
these spatial self-guided beams and estimate the power level required
for their experimental observation.
\end{abstract}

\begin{multicols}{2}
\narrowtext

Self-guided
optical beams (or spatial solitons) have attracted 
significant research
interest because they offer the possibility of all-optical switching and
controlling light by light (see, e.g.,~\cite{segev}). During the past few 
years it has been realised that quadratic nonlinearity is particularly 
attractive for potential practical realizations of all-optical switching, 
in that it not only supports stable solitons both in
planar waveguides and bulk media but also provids an ultra-fast
electronic nonlinear response (see, e.g.,~\cite{stegeman}).
The advantages of quadratic nonlinear materials are hampered in 
part by difficulties in obtaining close phase-velocity 
matching between interacting waves. One of the most 
effective ways to achieve such matching is the use of the so-called
quasi-phase-matching~(QPM) technique, in which large wave-vector mismatch 
between interacting waves is compensated for by periodic 
alternation of the sign of effective~$\chi^{(2)}$ coefficient.
This technique has been known since 1962, Ref.~\cite{jaa}, but only
in the past decade has technological progress put the QPM technique in the 
front line of modern nonlinear optics \cite{Fejer}.
In spite of all the theoretical and experimental progress achieved in the 
field of quadratic solitons, only solitons formed by  
waves with the same direction of propagation have been analysed so far 
(conventional {\em co-propagating} configuration; see, 
e.g.,~\cite{stegeman,wet_and_wild}). 
In a few works the advantages and implications of QPM technique
for this type of solitons have been analysed 
specifically~\cite{cbc}.
However, a parametric interaction between counterpropagating  
quasi-phase-matched waves in quadratic~[$\chi^{(2)}$] media 
is also possible. Corresponding analysis made for for non-soliton 
(plane wave interaction) case~\cite{yjdjbk} revealed certain advantages 
of a counterpropagating interaction system over conventional 
copropagating strategies. Moreover, recently a very similar so-called 
{\em backward} QPM configuration has been investigated
experimentally (see \cite{exper1}).
In this work we investigate the QPM counterpropagating scheme,
searching for solitary waves and investigating their stability. 

\begin{figure}
\vspace*{-2 mm}
\setlength{\epsfxsize}{8.3 cm}
\centerline{\epsfbox{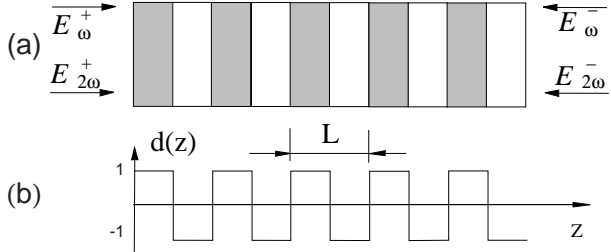}}
\caption{\protect\small Parametric interaction of counterpropagating waves. 
(a) Nonlinear grating is used to couple forward- (''+'') and
backward-propagating (''--'') waves; (b) square-wave function $d(z)$ 
approximates the periodic modulation of the effective second order nonlinear 
susceptibility coefficient.} 
\protect\label{grat}
\end{figure}

We consider the interaction between four
optical waves in a slab waveguide with an appropriate
nonlinear grating. 
Two forward-propagating waves, the fundamental at
frequency $\omega$ and with wave number $k_{\omega}$ and
the second harmonic ($2\omega,k_{2\omega}$), are coupled with
two backward-propagating ones, the fundamental ($\omega,-k_{\omega}$) and 
the second harmonic  ($2\omega,-k_{2\omega}$) [see Fig.~\ref{grat},~(a)]. 
Spatial modulation of nonlinear 
susceptibility along a crystal can be described in terms 
of square-wave function $d(z)$ [see Fig.~\ref{grat},~(b)].  
In this case the only nonzero matrix elements of the Fourier transform
of $d(z)$ are given by: $d_{l} = -2i/(\pi l)$, where $l=2m-1$, 
$m=\pm1, \pm2, \pm3\dots$ 
Following the method developed in Ref.~\cite{cbc} we can derive the 
corresponding normalised system of equations which has the following 
dimensionless form:
\begin{equation}
\label{general}
\begin{array}{r}
i\frac{\textstyle
\partial E_{\omega}^{+}}{\textstyle\partial z}+%
\frac{\textstyle\partial^2 E_{\omega}^+}{\textstyle\partial x^2}-%
\alpha_{+}E_{\omega}^{+}+A_{2\omega}E_{\omega}^{-*}=0, \vspace{1 mm}\\
-i\frac{\textstyle\partial E_{\omega}^{-}}{\textstyle\partial   z}+%
\frac{\textstyle\partial^2 E_{\omega}^{-}}{\textstyle\partial x^2}-%
\alpha_{-}E_{\omega}^{-}+A_{2\omega}E_{\omega}^{+*}=0,\vspace{1 mm} \\
i\sigma\frac{\textstyle\partial E_{2\omega}^{+}}{\textstyle\partial   z}+%
\frac{\textstyle\partial^2 E_{2\omega}^{+}}{\textstyle\partial x^2}-%
\sigma\beta_{+}E_{2\omega}^{+}+A_{\omega}^{+}E_{\omega}^{-}=0, \vspace{1 mm}\\
-i\sigma\frac{\textstyle\partial E_{2\omega}^{-}}{\textstyle\partial   z}+%
\frac{\textstyle\partial^2 E_{2\omega}^{-}}{\textstyle\partial x^2}%
-\sigma\beta_{-}E_{2\omega}^{-}+A_{\omega}^{-}E_{\omega}^{+}=0,
\end{array}\vspace*{- 2mm}
\end{equation}
where $A_{2\omega}\equiv(d_{-l}E_{2\omega}^{+}+d_{l} E_{2\omega}^{-})$,
$A_{\omega}^{+}\equiv 2d_{ l}E_{\omega}^{+}$,
$A_{\omega}^{-}\equiv 2d_{-l}E_{\omega}^{-}$,
$\beta_{+}\equiv\alpha_{+}-\alpha_{-}+\delta$, 
$\beta_{-}\equiv\alpha_{-}-\alpha_{+}+\delta$; 
$E^{\pm}_{\omega}(x,z),E^{\pm}_{2\omega}(x,z)$ are 
the envelopes of the fundamental 
wave and its second harmonic, respectively, 
sign ''+'' (''--'') corresponds to
the forward (backward) propagating wave;   
$x$ is the transverse coordinate
normalised on the width of the beam $r_0$; 
$z$ is the propagation distance
which is normalised on the diffraction length 
$z_d=k_{2 \omega}r_0^2$;
parameters $\alpha_+$ and $\alpha_-$ are nonlinear induced 
propagation constant shifts of the fundamental waves. Other two system
parameters, $\sigma \equiv k_{2\omega}/k_{\omega} = 2$ and 
$\delta \equiv (2\pi l/L-k_{2\omega}) z_d$ (where $L$ is 
the period of the nonlinear grating and $l$ is the order of QPM), are  
defined by the particular experimental setup. 
Note, that in contrast to Ref.~\cite{cbc} we have omitted all effective cubic terms 
in Eqs.~(\ref{general}). This is well justified  
for lower order QPM (for $l<15$) because in this case the cubic terms would 
become noticeable only when the light intensity exceeds the damage threshold for 
the typical nonlinear crystal. 

The system (\ref{general}) has the following family of power-like
integrals of motion:
\begin{equation} 
\begin{array}{c}
\label{inv}
Q(p_1,p_2) =
\int_{-\infty}^{+\infty} \{2p_1|E_{\omega}^{+}|^2 + 2p_2|E_{\omega}^{-}|^2+   \\
+\sigma(p_1-p_2)(|E_{2\omega}^{+}|^2 - |E_{2\omega}^{-}|^2) \}\,dx,
\end{array}
\end{equation} 
where $p_1,\: p_2$ are any real numbers. 
Using the asymptotic expansion technique 
and the method based on integrals of motion (see Ref. \cite{ktb}) one
can demonstrate that the stability threshold for 
the fundamental family of stationary  localised solutions of the 
system (\ref{general}) is given as:
\begin{equation}
\label{threshold}
\frac{\partial(Q_1,Q_2)}{\partial(\alpha_+,\alpha_-)}=
\frac{\partial Q_1}{\partial\alpha_+}
\frac{\partial Q_2}{\partial\alpha_-}-\frac{\partial Q_1}{\partial\alpha_-}
\frac{\partial Q_2}{\partial\alpha_+} = 0,
\end{equation}
where $Q_1$, $Q_2$ are any two
linearly independent invariants from the family 
(\ref{inv}), calculated for
the fundamental stationary  solitons, e.g. $Q_1 = Q(1,0)$ and
$Q_2 = Q(0,1)$. More elaborate analysis shows
that for the instability domains either of the two following conditions is satisfied 
in the vicinity of stability-instability boundaries:
\begin{eqnarray}
\label{lyapunov}
\frac{\partial Q_1}{\partial\alpha_{+}}
(\frac{\partial Q_1}{\partial\alpha_+}\frac{\partial Q_2}{\partial\alpha_-}-
 \frac{\partial Q_1}{\partial\alpha_-}\frac{\partial Q_2}{\partial\alpha_+}) < 0,  \nonumber \\
\frac{\partial Q_2}{\partial\alpha_-}
(\frac{\partial Q_1}{\partial\alpha_+}\frac{\partial Q_2}{\partial\alpha_-}-
 \frac{\partial Q_1}{\partial\alpha_-}\frac{\partial Q_2}{\partial\alpha_+}) < 0.\
\end{eqnarray} 
\normalsize

\begin{figure}
\vspace{-2 mm}
\setlength{\epsfxsize}{8.3 cm}
\centerline{\epsfbox{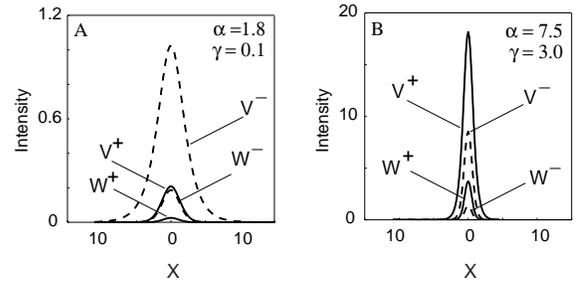}}
\caption{\protect\small
Examples of stationary solitons due to counterpropagating QPM
configuration. Profiles are calculated for the system (\protect\ref{num}).  
Solid curves correspond to the forward set of
waves, dashed lines~- to the backward-propagating ones.}
\label{profiles}
\end{figure}

At $\alpha_{+} = \alpha_{-}$ and $\delta=\sigma\alpha_{+}/2$
the system (\ref{general}) has 
an exact analytical stationary localised
solution in the form of $E^{\pm}_{\omega,s}(x)=
3\alpha_{+}/[4|d_l|{\rm cosh}^2(\sqrt{\alpha_{+}}x/2)]$,
$E^{\pm}_{2\omega,s}(x)=\mp iE^{\pm}_{\omega,s}(x)$.
For other values of the parameters, analytical
expressions cannot be found 
and numerics should be used.
For numerical analysis it is more convenient to renormalise 
the system (\ref{general}) reducing the number of parameters.
As a result we obtain the system
\begin{equation}
\begin{array}{r}
\label{num}
i\frac{\textstyle\partial V^{+}  }{\textstyle\partial Z  }+ 
\frac{\textstyle\partial^2 V^{+}}{\textstyle\partial X^2}- 
V^{+}+(W^{+}+W^{-})V^{-*}=0, \vspace{1 mm}          \\
-i\frac{\textstyle\partial V^{-}}{\textstyle\partial Z  }+ 
\frac{\textstyle\partial^2 V^{-}}{\textstyle\partial X^2}- 
\gamma V^-+(W^{+}+W^{-})V^{+*}=0, \vspace{1 mm} 	\\
i\sigma\frac{\textstyle\partial W^{+}}{\textstyle\partial Z  }+ 
\frac{\textstyle\partial^2 W^{+}}{\textstyle\partial X^2}- 
\sigma\beta_{W}^{+}W^{+}+2V^{+}V^{-}=0, \vspace{1 mm} 	\\
-i\sigma\frac{\textstyle\partial W^{-}}{\textstyle\partial Z  }+ 
\frac{\textstyle\partial^2 W^{-}}{\textstyle\partial X^2}- 
\sigma\beta_{W}^{-}W^{-}+2V^{+}V^{-}=0, 
\end{array}
\end{equation}
where $\beta_{W}^{+}\equiv1-\gamma+\alpha$,
$\beta_{W}^{-}\equiv\gamma-1+\alpha$. The connection between 
Eqs. (\ref{general}) and Eqs. (\ref{num})  
is given by scaling $E_{\omega}^{\pm}=\alpha_+V^{\pm}/|d_l|$, 
$\:E_{2\omega}^{\pm}=\mp i\alpha_+W^{\pm}/|d_l|$, 
$\:x=X/\sqrt{\alpha_+}$, $\:z=Z/\alpha_+$. System 
(\ref{num}) has only 
two parameters, defined as $\gamma\equiv\alpha_-/\alpha_+$,
$\:\alpha\equiv\delta/\alpha_+$. Stationary 
solitons of the system (\ref{num}) can be found numerically, e.g., using
the relaxation technique,
for all $\gamma > 0$, $\beta_{W}^{\pm} >0$. 
Some examples of stationary 
solitons of the system (\ref{num}) are shown in Fig.~\ref{profiles}.

Criteria (\ref{threshold}), (\ref{lyapunov}) can be rederived 
for the system (\ref{num}), and then used to calculate
the boundary of stability area in the $(\alpha,\gamma)$ plane. 
For the precise calculations of the invariants we use the tangential 
transformation of transverse soliton coordinate $X = \tan\pi\tilde{X}$,
Ref.\cite{fr}. Using this map we cover the infinite interval
in $X$ $(-\infty~<~X~<~\infty)$ by a finite one 
in $\tilde{X}$ $(-1/2~<~\tilde{X}~<~1/2)$.
\begin{figure}
\vspace*{-4 mm}
\setlength{\epsfxsize}{8.3 cm}
\centerline{\epsfbox{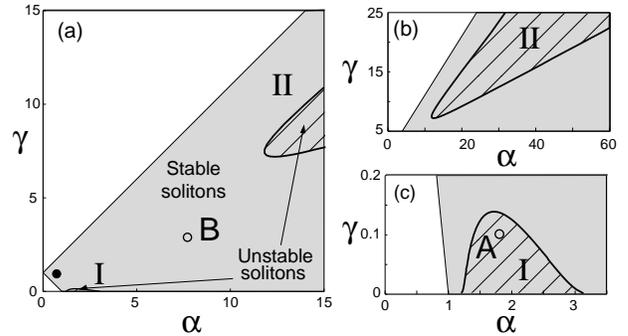}}
\caption{\protect\small (a) Existence and
stability diagram for the stationary solutions of the 
system (\protect\ref{num}). Solid curves are defined by the stability 
threshold condition (\protect\ref{threshold}); (b),~(c)~enlarged plots
of the unstable regions I and II. Soliton profiles calculated for the points 
A and B are shown in Fig.~\protect\ref{profiles}. Filled circle 
corresponds to the exact solution (see text).}
\protect\normalsize
\label{stab}
\end{figure}

\begin{figure}
\setlength{\epsfxsize}{8.3 cm}
\centerline{\epsfbox{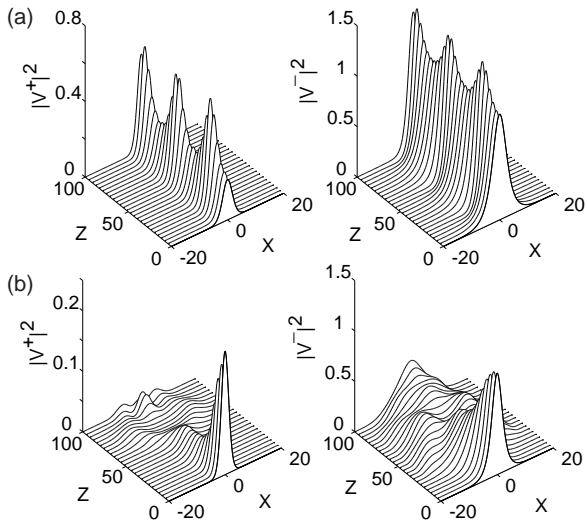}}
\caption{\protect\small Examples of unstable propagation of a slightly
perturbed stationary soliton shown in~Fig.~2,~A. 
Two typical kinds of instability 
correspond to slightly increased, (a), and 
reduced, (b), soliton amplitudes. Unstable soliton with a slightly increased
amplitude is clearly evolving to some stable state with an excited internal
mode (see, e.g., \protect\cite{intern}).}
\label{propagation}
\end{figure}

To confirm the validity of the stability 
results given by Eqs. (\ref{threshold}) and (\ref{lyapunov})  we 
analyse numerically an eigenvalue-eigenvector problem corresponding to 
Eqs. (\ref{num}) linearised about stationary solitons of interest.
In all analysed cases the theoretically predicted stability-instability
properties were confirmed numerically. However, we should note
that the theoretical approach which we use does not describe
so-called {\em oscillatory} instabilities (see, e.g.,  
\cite{oscil}) and, although we have not detected such 
instabilities numerically, further analysis is necessary to 
completely rule out their possibility.
The stability/instability domains
given by  criteria (\ref{threshold}) and (\ref{lyapunov}) are shown in Fig. 3.
Direct numerical modelling of Eqs. (\ref{num})
confirms the results of our stability 
analysis. Two examples of propagation of unstable solitons are presented in 
Fig.~\ref{propagation}. 

Optical power required for generation of counterpropagating solitons 
of a given beam width can be estimated by the method proposed in
Ref. \cite{avbysk}. After defining the soliton width $R_s$ as the 
maximum width at the half-maximum of the the second harmonic amplitude
$E_{2\omega}^{+}$ the conditional minimum $\tilde{Q}_{min}$ of the total power 
functional {\small $\tilde{Q}(\alpha,\gamma) = 
R_s^{3/2}\int_{-\infty}^{+\infty}\!\{|V^+|^2+|V^-|+2\sigma|W^+|^2+2\sigma|W^-|^2\}\,dX$} 
has to be found. Then minimal power density $I_{min}$ can be calculated as:
\begin{equation}
\label{min}
I_{min} = \tilde{Q}_{min}\frac{L_{eff}}{4\pi\chi^2_{bulk}r_0^4|d_l|^2},
\end{equation}
where nonlinear coefficient $\chi_{bulk}=(\omega^2/c^2)(2/c\epsilon_0n)^{1/2} \chi^{(2)}$
is expressed in $W^{-1/2}cm^{-1}$, $\chi^{(2)}$ is the effective
element of the second order susceptibility tensor, 
$L_{eff}$ is the width of a waveguide. For the first order QPM, i.e. $l=1$, 
and the values from Ref.~\cite{wet_and_wild}: $r_0 = 20 \mu m$, 
$\lambda= 2\pi c/\omega\simeq 1.064
\mu m$, refractive index $n\simeq 1.79$, $\chi^{(2)}\approx 6 pm/V$ and 
a waveguide with effective width $L_{eff}$ about $1\mu m$, we obtain
$I_{min}\approx 2.8 \; W/ \mu m$ ($\tilde{Q}_{min} \approx 103$ at 
$\alpha=3.15, \gamma=1.00$). 

Our analysis shows that the point of 
optimal generation is in the domain of stability and corresponds to  
$V^+ = V^-$ and $W^+ = W^-$, i.e.,
one can generate the whole four-wave soliton using only two seeded 
forward-propagating waves at one end of a crystal and
a mirror on  the  other [the use of mirror will also halve the generation 
threshold (\ref{min})]. Thus the solitons due to counterpropagating QPM, 
in principle, require less optical
power for an experimental observation in comparison with conventional
quadratic solitons, for which the corresponding value is 
$I_{min}\approx 3.4 \; W/ \mu m$ (see Ref.~\cite{avbysk}).
In practice lowering the generation threshold 
requires a very short (sub-micron) grating period to arrange for lower 
order QPM. This was not the case for 
the experiments \cite{exper1} where the grating period was about $3\mu m$
and  QPM order was high. However, experimental progress in 
quantum well technology (see, e.g., \cite{pavel}) may
make lower order counterpropagating QPM experimentally possible in the near 
future. 

In conclusion, we have demonstrated the existence of solitons that are 
due to counterpropagating QPM in quadratic media. We 
obtained an analytic criterion for the stability threshold for these
solitons and found a substantial region of stability with only two small  
regions of unstable solitons. We also discussed the
conditions for experimental observation of these novel solitons. 

The authors thank V.~V.~Steblina for helpful discussions and suggestions.
A.V.B. and R.A.S. acknowledge support from the 
Australian Research Council.

\end{multicols}
\end{document}